\documentclass[a4paper,preprint,amsmath,groupedaddress]{revtex4}

\usepackage{graphicx}
\usepackage{units}
\usepackage{natbib}
\usepackage{mathrsfs}
\usepackage{ulem}
\usepackage{color}

\begin{document}

\title{The role of metal ions in X-ray induced photochemistry}

\author{V. Stumpf}

\author{K. Gokhberg}
\email{kirill.gokhberg@pci.uni-heidelberg.de}

\author{L. S. Cederbaum}
\affiliation{Theoretische Chemie, Physikalisch-Chemisches Institut,
Universit\"at Heidelberg, Im Neuenheimer Feld 229, D-69120 Heidelberg,
Germany}

\date{\today}

\begin{abstract}

Metal ions play numerous important roles in biological systems being central to the function of biomolecules.
In this letter we show that the absorption of X-rays by these ions leads to a complicated chain of ultrafast
relaxation steps resulting in the complete degradation of their nearest environment. We conducted high quality 
{\it ab initio} studies on microsolvated Mg$^{2+}$ clusters demonstrating that ionisation of an 1s-electron of Mg 
leads to a complicated electronic cascade comprising both intra- and intermolecular steps and lasting only a few hundreds femtoseconds. 
The metal cation reverts to its original charge state at the end of the cascade, while
the nearest solvation shell becomes  multiply ionised and large concentrations of radical and slow electron species 
build up in the metal's vicinity. We conclude that such cascades involving metal ions are essential for understanding the 
radiation chemistry of solutions and radiation damage to metal containing biomolecules. 
    
\end{abstract}

\maketitle

Metals like Mg, Ca, Fe, Zn or Cu are essential for living organisms \cite{Bertini07}. They fulfil a number of 
important roles in biological systems, such as being instrumental for the catalytic activity of enzymes 
\cite{Bertini07,Andreini08} or ensuring structural stability of chromosomes \cite{Strick01,Wu10}.
In this report we investigate what is the role of these naturally occurring metal species in radiation damage or, 
more broadly, X-ray induced photochemistry of biological systems. The X-ray absorption spectroscopy of 
transition metal complexes in solutions offers a glimpse at the transformations accompanying the interaction of X-rays 
with the metal. The metal ion may become oxidised, reduced or even remain unaltered \cite{George08,Mesu06}; 
the particular outcome depends on the initial metal charge and the nature of ligands. Ample evidence from a different 
field, that of the X-ray crystallography, demonstrates that the radiation damage inflicted on a sample is highly
non-random \cite{Carugo05}. Importantly, in metallo-proteins the site containing the metal is one of the weak spots; 
it is often damaged at a dose much lower than the one at which the rest of the protein molecule is damaged 
\cite{Yano05}. It is common to consider the photoreduction of the metal ion as a manifestation of the damage 
\cite{George12}. This change in the metal's charge is thought to change the active site's geometry and 
correspondingly the diffraction image. 

In spite of accumulated knowledge mentioned above, details about the damage produced by the photoabsorption of an X-ray 
photon by the metal are sparse. Moreover, little is known about a sequence of events leading to this damage.
In what follows we present a new mechanism based on high-quality {\it ab initio} calculations,
which shows that the physico-chemical events started by the photoabsorption are concentrated on the 
metal species and its immediate vicinity causing thereby much damage to the molecules close to the metal.
We find that the damage caused following the absorption of an X-ray photon by the metal can also be substantial
in case the metal carries at the end of the process the same charge as before the photoabsorption.

Absorption of an X-ray photon by a metal predominantly removes a core electron and initiates a cascade of 
relaxation processes. For typical metals the first step is Auger decay during which electrons with characteristic 
energies are emitted by the metal species which in turn accumulates a positive charge. The emitted Auger electrons 
propagate through the system causing secondary ionisations and resulting in damaging lesions \cite{Howell08}. 
It is clear by now that in a biological medium further damaging relaxation processes are present. We expect the Auger 
decay to be accompanied and followed by ultrafast electronic decay processes involving the neighbouring molecules 
\cite{Gokhberg14}.

\begin{figure}[t]
 \begin{center}
  \includegraphics[scale=1.0]{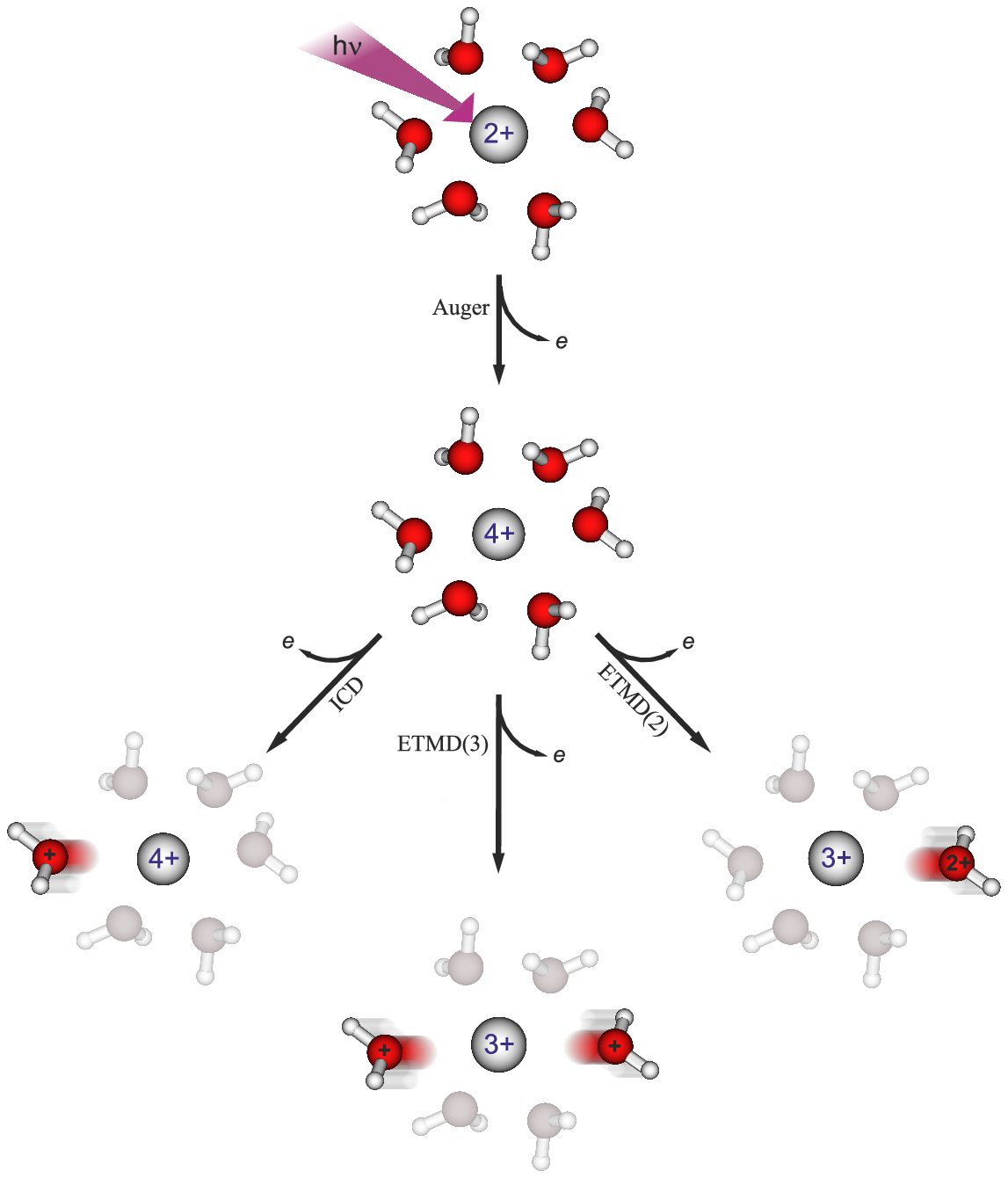}
  \caption{\small{Modes of electronic relaxation of the microsolvated Mg$^{4+}$ cation created by core ionisation of Mg$^{2+}$ (photoelectron not shown) and subsequent Auger
decay. Excited states of Mg$^{4+}$ transfer the excess energy to the environment ionising a water neighbour (ICD process). Low lying states of Mg$^{4+}$ accept an electron from a water neighbour, transferring the excess energy either to the same water molecule (ETMD(2) process) or to a different one (ETMD(3) process). All described modes of interatomic electronic relaxation are followed by Coulomb explosions.}}
  \label{Fig1}
 \end{center}
\end{figure}

These non-local processes come in two varieties (see Fig.\ref{Fig1}) - interatomic Coulombic decay (ICD) driven 
by energy transfer, and electron transfer mediated decay (ETMD) driven by charge transfer 
\cite{Cederbaum97,Zobeley01,Jahnke15}. 
In ICD electronically excited metal species (M) de-excites by ionising a neighbouring molecule (L)
\begin{equation}
\nonumber
M^{q+\ast}\cdot L_{n}\rightarrow M^{q+}\cdot L_{n-1} + L^{+} + e_{ICD}.
\label{eq1}
\end{equation} 
In ETMD a neighbour donates an electron to the metal species, while the excess energy is used to 
ionise the donor (ETMD(2)) or another molecule (ETMD(3))
\begin{subequations}
\begin{equation}
\nonumber
M^{q+}\cdot L_{n}\rightarrow M^{(q-1)+}\cdot L_{n-1} + L^{2+} + e_{ETMD(2)},
\label{eq2a}
\end{equation}    
\begin{equation}
\nonumber
M^{q+}\cdot L_{n}\rightarrow M^{(q-1)+}\cdot L_{n-2} + 2L^{+} + e_{ETMD(3)}.
\label{eq2b}
\end{equation}
\end{subequations}
Both ICD and ETMD lead to the ionisation of the nearest molecules and, in addition, to emitting a slow electron. However,
in ICD the medium is only singly ionised, while the charge on the metal remains unaltered. In ETMD the 
medium is doubly ionised and the charge on the metal is reduced by one. The rate of energy transfer 
processes falls off as an inverse power of the distance, while for the charge transfer processes the rate behaviour is
exponential \cite{Averbukh04,Zobeley01}. Therefore, previous work on the interatomic decay in rare-gas clusters 
showed that ICD proceeds on a timescale of tens of femtoseconds compared to the picosecond timescale of ETMD 
\cite{Kolorenc08}. As a result, ETMD is usually observed if the decaying state does not have enough excess energy 
to decay by ICD \cite{Forstel11,Sakai11}. This was specifically demonstrated to be the case for multiply charged ions 
produced in clusters by the Auger decay process \cite{Stumpf13}.

To study the complex chain of physico-chemical events following the X-ray photoabsorption by metal atoms
we considered a microsolvated cluster Mg$^{2+}$(H$_{2}$O)$_{6}$. It may serve as a model of Mg$^{2+}$ in solution
and previous experience showed that such cluster models work well \cite{Pokapanich11}.
Indeed, the Mg$^{2+}$ cation usually accommodates six water molecules in its first solvation shell.
The equilibrium Mg$^{2+}$-O distance in the cluster is 2.078 \AA~ and lies close to the corresponding value 
of 2.00-2.15 \AA~ in solution \cite{Ohtaki93}. Therefore, since the interatomic decay processes involve overwhelmingly 
the nearest neighbours, the decay rates and branching ratios should be also similar. More generally, the physico-chemical 
processes in the Mg$^{2+}$(H$_{2}$O)$_{6}$ cluster may serve as a paradigm applicable to other metal complexes.

We found (see Supplementary information for the computational details) that the binding energy of the magnesium 
1s-electron (K-edge) in Mg$^{2+}$(H$_{2}$O)$_{6}$ is 1317 eV. Removing this electron by an X-ray photon creates a highly 
energetic, electronically unstable trication. In the following we discuss the fate of this unstable trication. 
It predominantly decays by emitting an Auger electron and populating 
a number of Mg$^{4+}$(H$_{2}$O)$_{6}$ states. The states of Mg$^{4+}$ populated in the Auger decay are 
Mg$^{4+}(2p^{-2} [^1D,^1S])$ (63\%), Mg$^{4+}(2s^{-1}2p^{-1} [^1P])$ (21\%), Mg$^{4+}(2s^{-1}2p^{-1} [^3P])$ (9\%), 
and Mg$^{4+}(2s^{-2}[^1S])$ (7\%). All of them are unstable with respect to intermolecular 
electronic decay in the presence of water molecules. Since the excess energy in microsolvated
Mg$^{4+}(2p^{-2} [^1D,^1S])$ relative to the Mg$^{4+}(2p^{-2} [^3P])$ ground state of the cation is about 4-9 eV, 
these states cannot decay by ICD. However, they efficiently decay by both ETMD(2) and ETMD(3) pathways. The 
computed ETMD lifetime is 16.0 fs, which is much shorter than the picosecond lifetimes found in rare-gas clusters 
\cite{Kolorenc08}. Shorter ion - neighbour distances lead to larger orbital overlap greatly facilitating this electron 
transfer driven decay \cite{Zobeley01}. The ETMD(2) to ETMD(3) branching ratio is 1.0/1.6, therefore, 
both H$_{2}$O$^{+}$ and H$_{2}$O$^{2+}$ are efficiently produced
in the ionisation of the medium. The final products are either Mg$^{3+}(2p^{-1} [^2P])$(H$_{2}$O$^{2+}$)(H$_{2}$O)$_{5}$ 
or Mg$^{3+}(2p^{-1} [^2P])$(H$_{2}$O$^{+}$)$_{2}$(H$_{2}$O)$_{4}$; the charge on Mg decreases by one as the result of 
electron transfer from water (see Fig.\ref{Fig1}). The ETMD electrons have energies between 11 and 26 eV for ETMD(2) 
and between 23 and 40 eV for ETMD(3). The greater delocalisation of the positive charge in the final state of ETMD(3) 
results in faster emitted electrons.

Unlike the states considered above, microsolvated Mg$^{4+}(2s^{-1}2p^{-1} [^1P])$ possesses enough excess energy 
for the ionisation of the water molecules and, therefore, can decay by ICD. Its computed ICD lifetime is extremely short,
0.7 fs, showing that this interatomic decay occurs even faster than the already very fast local Auger decay on the metal 
(1.9 fs). We mention that the experimentally determined ICD lifetime of Mg$^{3+}(2s^{-1}) [^2S]$) in liquid water is 
1.0-1.5 fs \cite{Ohrwall10}, indicating that such extremely fast interatomic decay in hydrated metal ions is the rule 
rather than exception, due to the short metal-water distances and a large number of open ICD channels.
In this state ICD will also dominate ETMD by about an order of magnitude due to the higher efficiency of energy transfer.
The energies of the ICD electrons lie below 7 eV. Since ICD does not change the charge on Mg 
it produces Mg$^{4+}(2p^{-2})$(H$_{2}$O$^{+}$)(H$_{2}$O)$_{5}$ states. These are
the same states of  Mg$^{4+}$ we considered above, but now with H$_{2}$O$^{+}$ ion in its vicinity. Can they 
continue decaying by ETMD?
 
Model calculations with the water neighbour ionised via ICD replaced by a point charge (for details see 
Supplementary information) reveal that most ETMD(3) and ETMD(2) channels remain open. The ETMD lifetime grows from 
16.0 to 21.8 fs since only five neutral water neighbours are now available for the electronic decay. We also expect an 
interesting interplay between ETMD and fragmentation of the Mg$^{4+}(2p^{-2})$(H$_{2}$O$^{+}$)(H$_{2}$O)$_{5}$ cluster 
in a Coulomb explosion which typically proceeds on the femtosecond timescale \cite{Vendrell10}.

The Mg$^{4+}$ states Mg$^{4+}(2s^{-1}2p^{-1} [^3P])$ and Mg$^{4+}(2s^{-2}[^1S])$ which are weakly populated in
the Auger decay similarly exhibit complicated de-excitation pathways. They decay through a cascade of ICD and ETMD steps 
ultimately producing the Mg$^{3+}(2p^{-1} [^2P])$ cation. The lifetimes at all steps involved lie below 25 fs.

We saw that the electronic decay after the core ionisation of magnesium leads to the formation of Mg$^{3+}(2p^{-1} [^2P])$
species. We found that both ETMD(2) and ETMD(3) channels are open for Mg$^{3+}(2p^{-1} [^2P])$(H$_{2}$O)$_{6}$. The 
corresponding lifetime is 15.5 fs, while the emitted electrons have energies of 0-6 and 5-19 eV, respectively. Following 
this decay step the Mg$^{2+}$(H$_{2}$O$^{2+}$)(H$_{2}$O)$_{5}$ or Mg$^{2+}$(H$_{2}$O$^{+}$)$_2$(H$_{2}$O)$_{4}$ species 
are produced with the ratio of 1.0 to 1.2.

We wish to remark at this point the Mg$^{3+}(2p^{-1} [^2P])$ species can also be obtained directly from the initial core 
ionised Mg ion in the presence of water ligands. Thus, a 2p-electron of Mg may fill the 1s-vacancy and transfer its 
energy not to another 2p-electron, which would result in the Auger process on the metal, but to a valence electron on 
water ionising it in a core ICD-like process \cite{Pokapanich09,Pokapanich11,Slavicek14}. This decay leads
to population of Mg$^{3+}$(H$_{2}$O$^{+}$)(H$_{2}$O)$_{5}$ states. Since the positive charge in the latter is more 
delocalised than in the final states of the Auger decay, the electrons emitted in the core ICD-like process 
have larger energies (1205-1225 eV) compared to the Auger electrons (1111-1157 eV). What process will be more important? 
The Auger lifetime of the core ionised state was found to be 1.9 fs, while its ICD lifetime was 
57 fs. Therefore, the Auger decay takes place in 97 \% of Mg$^{4+}$(H$_{2}$O)$_{6}$ systems and core ICD-like only in 3\%. 
This branching ratio, however, can be tilted much more in favour of the interatomic process for other metal ions. 
Thus in the case of Ca$^{2+}$ the core ICD-like probability was found experimentally to be 10\% \cite{Ottosson12}.

\begin{figure}[t]
 \begin{center}
  \includegraphics[scale=0.35]{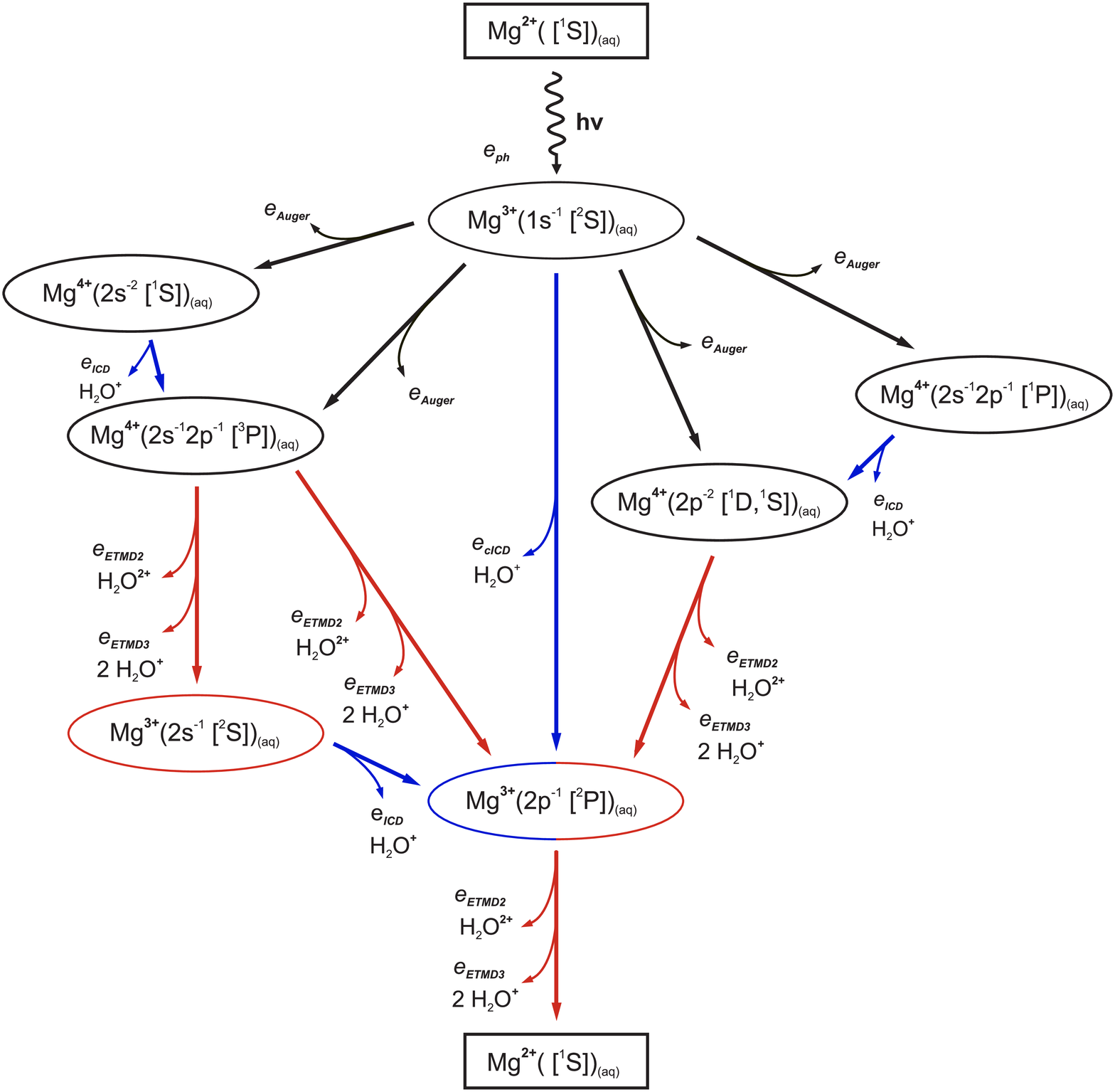}
  \caption{\small{Schematic description of the electronic decay processes taking place after the core ionisation of the 
Mg$^{2+}$ cation in aqueous solution. First, the intra-atomic Auger decay (black arrows) creates highly charged 
Mg$^{4+}$ cations. Then, the energetic ions lose their excess energy by ICD. The Mg$^{4+}$ ions with little excess energy 
undergo ETMD and become reduced. The cascade continues until all Mg$^{4+}$ and Mg$^{3+}$ ions revert to the original  
Mg$^{2+}$ state in a sequence of ultrafast, interatomic ICD (blue) and ETMD (red) steps. Estimated time for the 
complete cycle is 220 fs. Several reactive products such as electrons, water cations and dications are released mainly 
from the first coordination shell of Mg$^{2+}$ as indicated by the side arrows.}}
  \label{Fig2}
 \end{center}
\end{figure}

From the previous description of the decay processes involving different Mg cations it is clear that these 
disparate steps can be glued together in one continuous cascade. Its schematic representation is shown in Fig.\ref{Fig2},
where all steps discussed above can be found. Taking into account the decreasing 
number of neutral water neighbours available for interatomic decay at each consecutive step of the cascade we 
find that 90 percent of the core ionised states would cascade through to the final state within only 220 fs. 
Within this time window a large amount of Coulombic repulsion energy will be accumulated, due to the high metal charge 
and short distances to the ionised water neighbours. Its release is expected to result in a complicated fragmentation 
pattern, involving Coulomb explosions and molecular dissociation \cite{Pedersen13}. Note, that after only 10 fs 
already 28 \% of the core ionised states will undergo the Auger and the interatomic Coulombic decays. This sets off a 
Coulomb explosion leading to a modification of molecular geometry in the vicinity of the metal ion already at that short 
timescale.

All the individual decay steps we discussed above for the microsolvated clusters will occur also in solution. 
The presence of additional solvation shells will further increase the efficiency of ICD, but their overall impact 
will be moderate, mainly because of their larger distance to the metal ion \cite{Averbukh04,Fasshauer14}.
The effect of the polarisable medium will be manifested in stabilising of the positively charged ions at different 
stages of the cascade. Therefore, the energies of the emitted electrons will be somewhat different from the cluster's case.

This cascade may be viewed from two different angles: the one of the metal cation, and the other one of the nearest 
neighbours or ligands. As we mentioned above the core ionisation of Mg$^{2+}$ creates a highly metastable ion which  
first decays mostly in the Auger process. Even after the local Auger processes are over, the interatomic decay keeps going on. 
If for a specific electronic state of the metal both ICD and ETMD channels are available, ICD will be predominant.  
It will lead to the electronic relaxation of the metal ion without changing its charge. An ETMD step
will follow reducing the charge by one. This is illustrated by the decay of Mg$^{4+}(2s^{-1}2p^{-1} [^1P])$ first by ICD 
to Mg$^{4+}(2p^{-2})$ states and further by ETMD to Mg$^{3+}(2p^{-1} [^2P])$. Even if the ion does not have 
excess energy and ICD is not possible there can still be available ETMD channels. Therefore, the cascade of the 
interatomic processes goes on and the ion is being reduced until no interatomic electronic decay is allowed. 
For our system it leads to the surprising result that Mg ends up in the same electronic state it was in initially.

From the point of view of the nearest neighbours the probability of their damage by the $\approx$1150 eV Auger electrons 
is small. However, they are directly damaged in the interatomic processes becoming either singly or doubly ionised.
In the case of water solutions the single ionisation will lead to proton transfer reaction producing HO$\cdot$ radical,
while doubly ionised water is expected to produce atomic oxygen \cite{Tavernelli08}. 
The hydroxyl radical is highly reactive and causes oxidative damage to other molecules present in the solution 
(see e.g. \cite{Oneill01}). Atomic oxygen is reactive as well and is admittedly a source of further damaging species 
such as H$_2$O$_2$ \cite{Gaigeot07}. In addition to such direct damage to the nearest 
neighbours and the production of radicals, the interatomic processes result in the emission of electrons
having energies $<$40 eV. Such electrons can be resonantly captured by the molecules 
in the near environment initiating efficient bond breaking reactions \cite{Alizadeh15}. In total, 
the interatomic decay processes massively degrade the molecules in the immediate vicinity of the metal species 
through multiple ionisations releasing in their course both reactive electrons and radicals. 
If we are to neglect the interatomic processes, the decay of the 1s vacancy on Mg$^{2+}$ would result in one (Auger) electron. 
Taking them into account and counting the damaging particles released in the complete cascade 
one would obtain on average 2.4 slow electrons in addition to a fast Auger electron and 4.3 radicals per each 1s vacancy.

In the decay cascade presented above the Mg ion which absorbs the X-ray photon reverts back to its initial electronic 
state within few hundred femtoseconds. At the same time the nearest environment of this ion is multiply ionised 
implying the production of a large number of reactive particles (radicals and slow electrons) 
at the metal's location. This shows that no change of the metal ion's charge on X-ray irradiation is by no means 
equal to no damage done. On the ground of our detailed findings one may anticipate that the similar cascades generally 
hold also for other metal ions independently whether the ion reverts back to its initial charge or photoreduction 
is observed. Clearly, these cascades will be accompanied by extensive damage to the surrounding molecules. 
In addition, strategic positioning of metal ions in biomolecules makes such damage particularly disruptive for the 
functions these molecules fulfil. Experimental evidence is available that irradiation of DNA molecules complexed with 
Ca$^{2+}$ by X-rays at energies below and above the Ca K-edge showed about 30\% enhancement in the induction of double 
strand breaks when the photon energy was increased across the K-edge \cite{Takakura96}. Similarly, it was demonstrated 
that iron containing metallo-enzymes were more efficiently deactivated when irradiated with the X-rays above the iron's 
K-edge \cite{Jawad86}. We hope that the interatomic electronic cascades elucidated in this report will be useful in 
understanding the X-ray induced photochemistry and radiation damage of metal containing biomolecules.

\section{Methods}

All computations were done by \textit{ab initio} electronic structure methods, using triple-$\zeta$ level Dunning basis 
sets \cite{Dunning89,Woon94,Woon95,EMSL}. The only exception was the calculation of Auger decay rates and product 
populations, where quintuple-$\zeta$ level basis sets were required. For the basis set details see the Supplementary 
information.

The geometry of the Mg$^{2+}$(H$_2$O)$_6$ cluster was obtained by symmetry constrained optimisation, relying on the 
M{\o}ller-Plesset second order perturbation theory (MP2) implemented in the MOLPRO 2010.1 quantum chemistry package 
\cite{MOLPRO2010}. The T$_h$ symmetry of the cluster was chosen according to Feller et al. and represents the global 
minimum geometry \cite{Glendening96}. All electronic decay processes were studied at this cluster geometry. 

To identify open decay channels, the energies of the involved tricationic and tetracationic electronic states were 
calculated by the Algebraic Diagrammatic Construction (ADC(2)) scheme for the one-particle and two-particle propagator, 
respectively \cite{Trofimov05,Tarantelli06,Velkov11}. The latter allow the calculation of single and 
double ionisation potentials relatively to the Mg$^{2+}$(H$_2$O)$_n$ electronic ground state introducing a perturbational 
expansion scheme of the propagator complete up to second order. The electron integrals and molecular orbitals serving as 
input for the propagator based computations were calculated by MOLCAS 7.4 software package \cite{MOLCAS74}. The energies 
of pentacationic states were obtained by multi-reference configuration interaction method including single and double 
excitations (DIRECT-CI)\cite{Saunders83} implemented in the GAMESS-UK 8.0 quantum chemistry package \cite{gamess_uk}. 
The reference space was built up starting with the  Mg$^{2+}$(H$_2$O)$_n$ ground state Hartree-Fock determinant. 
The final states of ETMD were constructed by selecting all configurations having a hole in a 2s- or 2p-orbital of Mg 
plus two additional holes in the valence orbitals localised on water molecules. The final states of ICD were 
constructed by selecting all configurations having two holes in 2s- 2p-orbitals of Mg plus an additional hole in 
the valence orbitals localised on water molecules.

The total and partial electronic decay widths were calculated by means of Fano-ADC-Stieltjes method 
\cite{Averbukh05, Kolorenc08}. Due to numerical limitations, the decay widths for the Mg$^{4+}$(H$_2$O)$_6$ states were 
determined in an additive approximation \cite{Mueller05,Kryzhevoi07}. For details of this approximation and calculation 
of the partial decay widths see the Supplementary information.

\end{document}


\begin{center}
\textbf{Supplementary information}
\end{center}

\title{The role of metal ions in X-ray induced photochemistry}

\author{V. Stumpf}

\author{K. Gokhberg}
\email{kirill.gokhberg@pci.uni-heidelberg.de}

\author{L. S. Cederbaum}
\affiliation{Theoretische Chemie, Physikalisch-Chemisches Institut,
Universit\"at Heidelberg, Im Neuenheimer Feld 229, D-69120 Heidelberg,
Germany}

\maketitle

\section{Identification of the electronic states involved in the decay cascade}

Core ionisation of the Mg$^{2+}$(H$_2$O)$_6$ cluster creates a highly energetic magnesium trication which decays by the Auger mechanism and accumulates positive charge. A series of interatomic electronic decay steps leads to the reduction of magnesium to the original 
Mg$^{2+}$ state. The following steps of electronic decay were considered in this work:

\begin{center}
  \begin{align}
   \begin{split}
  Mg^{3+}(1s^{-1})(H_{2}O)_{6} & \xrightarrow{Auger} Mg^{4+}(H_{2}O)_{6} \\
  Mg^{3+}(1s^{-1})(H_{2}O)_{6} & \xrightarrow{Core~ICD-like} Mg^{3+}(H_{2}O^+)(H_{2}O)_{5}
    \end{split}
   \label{S1}
  \end{align}
\end{center}

\begin{center}
  \begin{align}
   \begin{split}
  Mg^{4+}(H_{2}O)_{6} & \xrightarrow{ICD} Mg^{4+}(H_{2}O^+)(H_{2}O)_{5} \\
  Mg^{4+}(H_{2}O)_{6} & \xrightarrow{ETMD(2)} Mg^{3+}(H_{2}O^{2+})(H_{2}O)_{5} \\
  Mg^{4+}(H_{2}O)_{6} & \xrightarrow{ETMD(3)} Mg^{3+}(H_{2}O^+)_2(H_{2}O)_{4}
   \end{split}
   \label{S2}
  \end{align}
\end{center}

\begin{center}
  \begin{align}
   \begin{split}
  Mg^{3+}(H_{2}O)_{6} & \xrightarrow{ICD} Mg^{3+}(H_{2}O^+)(H_{2}O)_{5} \\  
  Mg^{3+}(H_{2}O)_{6} & \xrightarrow{ETMD(2)} Mg^{2+}(H_{2}O^{2+})(H_{2}O)_{5} \\
  Mg^{3+}(H_{2}O)_{6} & \xrightarrow{ETMD(3)} Mg^{2+}(H_{2}O^+)_2(H_{2}O)_{4}.
   \end{split}
   \label{S3}
  \end{align}
\end{center}

The involved electronic states comprise tricationic, tetracationic and pentacationic states. For these states the energies had to be computed in order to determine open decay channels. All electronic structure calculations were carried out at the equilibrium geometry of the Mg$^{2+}$(H$_2$O)$_6$ cluster using the ground electronic state of Mg$^{2+}$(H$_2$O)$_6$ as the reference state. The initial Mg$^{2+}$(H$_2$O)$_6$ state energy was determined by the M{\o}ller-Plesset second order perturbation theory (MP2).

We start by considering the tricationic states, which appear as initial states in Eqs. (\ref{S1},\ref{S3}). Their energies were obtained by the Algebraic Diagrammatic Construction (ADC(2)) method: a propagator approach based on perturbational expansion of the Green's function complete up to second order. The ADC(2)-extended scheme was used throughout \cite{Trofimov05}. The energies of the tetracationic states were calculated by the ADC(2) method for two-particle propagator \cite{Schirmer84, Tarantelli06}.

The ETMD and ICD of high lying tetracationic states lead to pentacationic states (Eq. (\ref{S2})). For these states no propagator approach was available. Therefore, a multi-reference configuration interaction method including single and double excitations was used. The configuration interaction method is computationally demanding and not size extensive. These drawbacks limited the size of the model system to the metal cation and two water neighbours. 
The energy difference between the initial tetracationic and the final pentacationic states determines the kinetic energy of the ICD and ETMD electrons. 
To estimate the error introduced into the calculation of this energy difference by retaining only two water ligands instead of six we analysed the decay of the tricationic states (Eq. (\ref{S3})). The corresponding ETMD electron energies revealed a decrease of 1-2 eV when $n$ was increased from 2 to 6. For the energies of ICD electrons an opposite trend is observed: they grow by 3-4 eV as $n$ increases from 2 to 6. The energies of the ETMD electrons for Mg$^{4+}$(H$_2$O)$_2$ were found to lie above 10 eV, thus all ETMD channels open in Mg$^{4+}$(H$_2$O)$_2$ are expected to remain open in Mg$^{4+}$(H$_2$O)$_6$, while the ETMD electron energies are expected to be
shifted by few eV to lower values. The ICD electrons, lying below 7 eV for Mg$^{4+}$(H$_2$O)$_2$, are expected to be shifted to higher values.

\section{Determination of the widths of electronic decay}

In order to discuss the timescales on which the electronic decay processes occur, the corresponding widths were calculated. For this purpose we used the Fano-Stieltjes-ADC approach. Within this \textit{ab initio} approach the bound and continuum parts of the decaying state are constructed by means of the ADC(2)-extended scheme for the one-particle and two-particle propagator, respectively \cite{Averbukh05, Kolorenc08}. The renormalisation of the pseudocontinuum generated by the ADC scheme in an $L^2$ basis is done by the Stieltjes imaging technique \cite{Reinhardt79}. The partial decay widths were estimated by applying projector operators to the pseudocontinuum component of the wave function. The projector operators were defined by a channel or a group of channels of interest (e.g. all ETMD(2) channels). 

The limitation of the Fano-Stieltjes-ADC method is given by the size of the electronic interaction matrices to be diagonalised for the construction of the bound and pseudocontinuum components of the resonance wave function. For example, the interaction matrices of the tetracationic decaying states are essentially spanned by the large $3h1p$ block, making a full \textit{ab initio} computation of decay widths unfeasible for bigger systems. However, as was demonstrated in previous studies on ETMD and ICD, a good approximation to the electronic decay widths can be obtained by adding up decay widths calculated for subunits of a system \cite{Mueller05,Kryzhevoi07}.
\begin{figure}[h]
    \begin{center}
\includegraphics[scale=0.33]{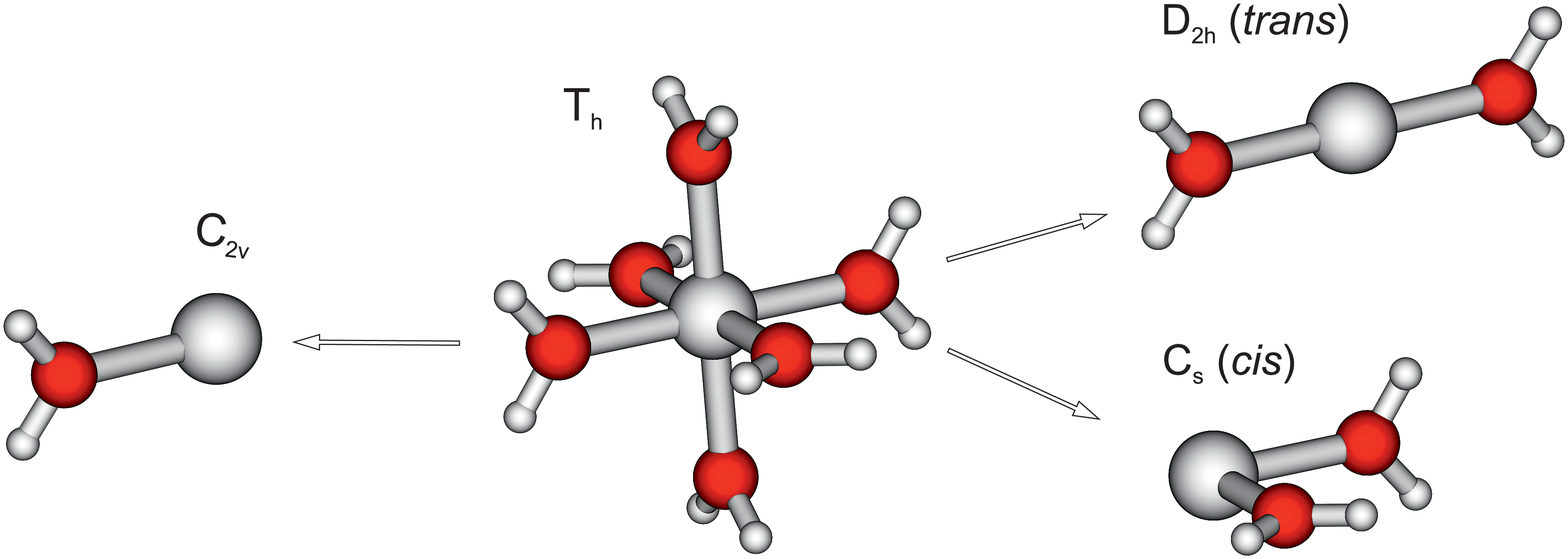}
\caption{Division of the Mg$^{q+}$(H$_2$O)$_6$ cluster into subunits required for the calculation of the decay widths by the additive approach. 
Note: the symmetry lowering may mix the states belonging to the different irreducible representations of the T$_h$ group of the Mg$^{q+}$(H$_2$O)$_6$ cluster.}
 \label{Fig1}
\end{center}
\end{figure}

The approximation can be exemplified for the total ICD width in a cluster with $N$ water neighbours:

\begin{equation}
\Gamma^{ICD} = \sum\limits_{i=1}^{N} \Gamma^{ICD}_{i},
\end{equation}

where $\Gamma^{ICD}_{i}$ is the ICD width due to the i-th water neighbour (see Fig.\ref{Fig1}). The total ETMD(2) width is obtained in a similar way:

\begin{equation}
\Gamma^{(2)} = \sum\limits_{i=1}^{N} \Gamma^{(2)}_{i}.
\end{equation}

For the determination of the total ETMD(3) width the system is divided into subunits containing the metal ion and pairs of neighbours $ij$:

\begin{equation}
\Gamma^{(3)} = \sum\limits_{i<j} \Gamma^{(3)}_{ij}.
\end{equation}

The pairs in the case of the Mg$^{q+}$(H$_2$O)$_6$ clusters (see Fig. \ref{Fig1}) can further be divided into $trans$ and $cis$ type pairs:

\begin{equation}
\Gamma^{(3)} = \sum \Gamma^{(3)}_{ij}(trans) + \sum \Gamma^{(3)}_{ij}(cis).
\end{equation}

The Mg$^{3+}$(H$_2$O)$_6$, Mg$^{4+}$(H$_2$O)$_6$ electronic states correlate with the atomic multiplets such as Mg$^{4+}(2p^{-2} [^1D])$. The multiplets are weakly split (\textless 0.1 eV) in the ligand field of the water molecules. Moreover, we assume that they are equally populated in the electronic decay. To simplify the calculations while retaining their accuracy we computed the interatomic decay widths averaged over such multiplets:

\begin{equation}
\overline{\Gamma} = \frac{1}{l} \sum_m^l{\Gamma_m},
\end{equation}

where $l$ denotes the multiplicity and $m$ an individual member of the multiplet. A total decay width of Mg$^{q+}$(H$_2$O)$_6$ is calculated by the summation of the multiplet averaged decay widths due to the subunits:

\begin{equation}
\overline{\Gamma} = \sum\limits_{i} \overline{\Gamma}_i = \sum\limits_{i}  \frac{1}{l} \sum_m^l{(\Gamma_m)_i}.
\end{equation}

We compared the results of a full \textit{ab initio} calculation with the values determined by the additive approach for the Mg$^{3+}(2p^{-1} [^2P])$(H$_2$O)$_6$ states. The corresponding multiplet averaged ETMD(2) width is given by:

\begin{equation}
\overline{\Gamma}^{(2)} = \sum\limits_{i=1}^{6}  \frac{1}{3} \sum_{m=1}^3{(\Gamma_m^{(2)})_i} = \frac{6}     {3}\sum_{m=1}^3{\Gamma_m^{(2)}}.
\end{equation}

For the ETMD(3) width we obtain:

\begin{equation}
\overline{\Gamma}^{(3)} = \frac{3}{3}\sum_{m=1}^3{\Gamma_m^{(3)}(trans)} +\frac{12}{3}\sum_{m=1}^3{\Gamma_m^{(3)}(cis)}.  
\end{equation}

The ETMD lifetime of Mg$^{3+}(2p^{-1} [^2P])$(H$_2$O)$_6$ obtained by the additive approach (15.6 fs) lies close to the value obtained by the full \textit{ab initio} calculation (17.3 fs). The branching ratios of ETMD(3) to ETMD(2) product populations are 1.50 and 1.23, respectively. Therefore, we conclude that the approximate method is able to reproduce the total decay width and the branching ratios fairly well.

The additive approach was utilised to compute the decay rates of the Mg$^{4+}$(H$_2$O)$_6$ states. We assumed that all decay channels which are open in a subunit with a single neighbour ($n=1$) or a pair of neighbours ($n=2$) will remain open in the larger ($n=6$) system (see the discussion in the previous section). In the Table \ref{Tab1} the lifetimes of the states populated in the Auger decay of core ionised Mg$^{3+}$(H$_{2}$O)$_{6}$ are given. The ICD rates are of an order of magnitude higher than the ETMD rates. The states decaying by ETMD demonstrate comparable ETMD(2) and ETMD(3) efficiencies, which can be explained by the size of the system. Although ETMD(2) is faster in a subsystem with two neighbours, the quadratic scaling of the number of ETMD(3) channels with the number of neighbours $n$ strongly enhances ETMD(3) for $n$=6.
Compared to the Mg$^{4+}(2p^{-2} [^1D,^1S])$ states, the higher lying Mg$^{4+}(2s^{-1}2p^{-1} [^3P])$ state exhibits a more complex
ETMD pattern. Both 2s and 2p holes on magnesium may act as the electron acceptor, producing either Mg$^{3+}(2p^{-1} [^2P])$ or
Mg$^{3+}(2s^{-1} [^2S])$ cations. The electron transfer into the 2p hole is favoured because of the greater overlap with orbitals of water.

\begin{table}
\caption{Decay lifetimes of cations produced by Auger process in the core
ionised Mg$^{3+}$(H$_{2}$O)$_{6}$  cluster. Minor channels, contributing $<10$ percent to the product population,
are not shown.}
\begin{tabular}{ c c c c c}\\
 Decaying state & Lifetime [fs]& ETMD(3) & ETMD(2) & ICD \\
 \hline
 $2p^{-2} [^1D]$ & 15.6 & 0.65 & 0.35 &  \\\
 $2p^{-2} [^1S]$ & 18.2 & 0.57 & 0.43 &  \\
 $2s^{-1}2p^{-1} [^3P]$ & 17.5 & 0.51 & 0.49 &  \\
 $2s^{-1}2p^{-1} [^1P]$ & 0.7  &      &                 & 1.0 \\
 $2s^{-2} [^1S]$        & 0.9  &      &                 & 1.0 \\
\end{tabular}
\label{Tab1}
\end{table}

The additive approach was furthemore applied to calculate the decay rate of the core ICD-like processes in the Mg$^{3+}$(H$_{2}$O)$_{6}$ cluster. The core ionised Mg$^{3+}(1s^{-1} [^2S])$(H$_{2}$O)$_{6}$ state decays not only locally by the Auger mechanism but is also coupled to a very large number of final states of interatomic electronic decay (e.g. core ICD-like processes). Therefore, an explicit computation of total and partial rates of decay for this state was not feasible. The rate of the core ICD-like processes was calculated for a subsystem with one water neighbour and extrapolated to $n=6$. The Auger rate in the Mg$^{2+}$(H$_2$O)$_6$ cluster was expected to be only weakly modified compared to the atomic value, which was computed by the Fano-Stieltjes-ADC method.

\section{Cascade of electronic decay processes}

The separate electronic decay steps considered in this work can be combined into a decay cascade. The accumulation of the positive charge in the course of the decay cascade will have an impact on the electron energies and decay lifetimes. In general, the latter are expected to increase due to the fact that less ionisable water neighbours are available after each step of the cascade. The additive approach allows for a recalculation of the decay lifetime for the given number of ionisable neighbours $n$ and a given decay step. Since any electronic decay step leads to the increase of the positive charge of the cluster by one, the final states of the decay will be destabilised relative to the initial ones. Thus, the number of available decay channels will further decrease and the electron energies will be shifted to lower values.

We took this effect into account replacing ionised neighbours by point charges in the \textit{ab initio} calculations of the energies of the final and initial states of decay. The point charge model is a crude approximation, however, it contains the leading term in the multipole expansion of the charge distribution on the ionised water molecules. Thus, the main contribution of the electrostatic interaction with the ionised water neighbours is included.
Such calculations show that the last step of the most prominent pathway in the decay cascade:
\begin{align}
 \begin{split}
Mg^{3+}(1s^{-1} [^2S])(H_{2}O)_{6} & \xrightarrow{Auger} Mg^{4+}(2p^{-2} [^1D])(H_{2}O)_{6} \\ & \xrightarrow{ETMD(3)} Mg^{3+}(2p^{-1} [^2P])(H_{2}O^+)_{2}(H_{2}O)_{4}\\  & \xrightarrow{ETMD(3)} Mg^{2+}[^1S](H_{2}O^+)_{4}(H_{2}O)_{2}
 \end{split}
\end{align}
is energetically possible without removing ionised water molecules produced in the first ETMD(3) step. 

Using the point charge model to obtain the decay widths of the individual steps (see Fig. 2 in the main text), we estimated the time necessary for the completion of the cascade in the Mg$^{3+}(1s^{-1})$(H$_{2}$O)$_{6}$ cluster. An Mg$^{2+}$ population of 0.9 is found at 220 fs, whereas the true time of the completion is expected to be shorter. Fast repulsive nuclear dynamics following ICD and ETMD, which we did not take into account, will efficiently remove the ionised neighbours and open closed decay channels.

\section{Basis sets for electronic structure calculations}

A characteristic feature of the electronic ground states of the Mg$^{2+}$(H$_2$O)$_n$ clusters as well as of the decaying states is the positive charge at the metal site. In this case cc-pCVTZ basis set on magnesium and  daug-cc-pVTZ basis set on oxygen and hydrogen atoms were utilised for geometry optimisations and calculations of energies of the decaying states \cite{EMSL,Woon94}. The calculation of the core ionisation potential of Mg$^{2+}$(H$_2$O)$_6$ required an uncontraction of the atomic basis set on magnesium. In the final states of the electronic decay additional positive charge on the neighbouring water molecules is created, therefore cc-pCVTZ basis set for oxygen and cc-pVTZ for hydrogen atoms were chosen \cite{Woon95,Dunning89}.

Decay rate calculations required an augmentation of the standard atomic basis sets by Kaufmann-Baumeister-Jungen (KBJ) continuumlike functions to improve the description of the pseudocontinuum \cite{Kaufmann89}. In particular, cc-pCVTZ (magnesium and oxygen) and cc-pVTZ atomic basis set (hydrogen) were augmented by 1\textit{s}1\textit{p}1\textit{d}1\textit{f} KBJ basis functions on magnesium, 4\textit{s}4\textit{p}4\textit{d}1\textit{f} on oxygen and 2\textit{s}2\textit{p}2\textit{d} on hydrogen. Additionally, 2\textit{s}2\textit{p}2\textit{d} KBJ basis functions were located between the magnesium and oxygen atoms. An uncontraction of the atomic cc-pCV5Z (magnesium, oxygen) and cc-pV5Z (hydrogen) basis sets was done for the computation of the Auger rate, in order to describe the fast outgoing Auger electron \cite{EMSL,Woon95}.